\pdfoutput=1
\documentclass[10pt,twoside]{predoc2}

\usepackage{bm,mathrsfs}
\usepackage{IEEEtrantools}
\usepackage{subcaption,array,booktabs}
\usepackage[flushleft]{threeparttable}
\usepackage{graphicx}
\usepackage{algorithmicx}
\usepackage{algpseudocode}
\usepackage{algorithm}
\usepackage[inline]{enumitem}
\usepackage{setspace}
\usepackage[title]{appendix}
\usepackage{float}
\usepackage[dvipsnames]{xcolor}
\usepackage{tikz}
\usepackage{tikz-qtree}
\usetikzlibrary{trees}
\newfloat{algorithm}{tp}{toa}

\definecolor{projpredcolour}{HTML}{009E73}
\definecolor{autocolour}{HTML}{E69F00}
\definecolor{refcolour}{HTML}{F0E442}
\definecolor{badcolour}{RGB}{213,94,0}
 
\numberwithin{equation}{section}

\newcommand{\pkg}[1]{{\fontseries{b}\selectfont #1}}
\let\proglang=\textsf
\DeclareMathOperator*{\argmin}{arg\,min}
\DeclareMathOperator{\elpd}{elpd}
\DeclareMathOperator{\arma}{ARMA}
\DeclareMathOperator{\ar}{AR}
\DeclareMathOperator{\ma}{MA}
\DeclareMathOperator{\auto}{auto}
\DeclareMathOperator{\normal}{N}
\DeclareMathOperator{\student}{Student-}

\title{Bayesian order identification of ARMA models with projection predictive inference}
\author{Yann McLatchie}
\author{Asael Alonzo Matamoros}
\author{David Kohns}
\author{Aki Vehtari}
\affil{Department of Computer Science, Aalto University, Espoo, Finland}

\makeatletter
\def\runauthor{McLatchie, Matamoros, Kohns, and Vehtari}
\def\runtitle{Projection predictive ARMA order identification}
\makeatother

\begin{document}
\maketitle
\thispagestyle{empty}
\begin{abstract}
    Auto-regressive moving-average (ARMA) models are ubiquitous forecasting tools. Parsimony in such models is highly valued for their interpretability and computational tractability, and as such the identification of model orders remains a fundamental task. We propose a novel method of ARMA order identification through projection predictive inference, which benefits from improved stability through the use of a reference model. The procedure consists of two steps: in the first, the practitioner incorporates their understanding of underlying data-generating process into a reference model, which we latterly project onto possibly parsimonious submodels. These submodels are optimally inferred to best replicate the predictive performance of the reference model. We further propose a search heuristic amenable to the ARMA framework. We show that the submodels selected by our procedure exhibit predictive performance at least as good as those chosen by AIC over simulated and real-data experiments, and in some cases out-perform the latter. Finally we show that our procedure is robust to noise, and scales well to larger data.
\end{abstract}
\section{Introduction}
Since their introduction by \citet{boxjen:1970}, auto-regressive moving-average models (ARMA) have become ubiquitous forecasting tools. This is thanks to their high predictive power, ease of implementation, and intuitive interpretation. Their use is often argued from a theoretical perspective in time series analysis in that any stationary time series process can be represented by an infinite order moving-average (MA) model, according to the Wold decomposition theorem \citep{wold1938study}. Such MA models can then be arbitrarily well approximated by appropriate finite order ARMA models. From an empirical perspective, they play a pivotal role in the social sciences (among many others), and particularly economics. Indeed, macroeconomic time series are often sums of underlying sub-time series (e.g. disaggregated inflation items feeding into headline inflation) and thus ARMA models naturally arise due to the sum of auto-regressive time series theorem \citep{granger1976time}. Further, many structural economic models have a moving averaging representation \citep{giacomini2013relationship}, and ARMA models have been shown to have competitive forecast performance for aggregate economic time series \citep{stock2007has, koop2013forecasting, chan_moving_2013, zhang2020stochastic}.

While we don't claim that ARMA models are better than any other in terms of their predictive performance, they can provide the statistician with information on correlation structures and baseline predictive performance. When building more complex models, such baseline models can operate as sanity checks and are commonly employed in statistical analyses due to their effectiveness given their relative simplicity.

Indeed, parsimony in these models is highly valued, not least due to the fact that latent MA components often result in difficult likelihoods with increasing order \citep{chib1994bayes,ives2010analysis,chan_moving_2013}. Thus one primary goal when implementing ARMA models is to identify small orders capable of good predictive performance. One popular approach consists of fitting many such models with different orders and selecting the best one according to some criterion, most famously the Akaike information criterion \citep[AIC;][]{autoarima}. This approach, however, has been known to not always select parsimonious models, and may select models with unexpectedly ragged temporal structures. 

In this paper we propose a new method of selecting ARMA and seasonal ARMA orders motivated from fully Bayesian decision theory. The proposed methodology contributes to projection predictive inference, originally defined by \citet{goutis1998} and later developed by \citet{dupuis2003}, \citet{piironen2016}, and \citet{piironen2018} by making it amenable to detecting relevant ARMA and seasonal ARMA orders from a predictive perspective. The method follows a two step procedure in which the modeller first specifies a possibly large model, which incorporates all relevant knowledge about the underlying data generating process of the data, and passes posterior predictive checks \citep{Pavone2020,bayesianworkflow,gabry_visualization_2019}.\footnote{With provision of a reference model, the model space is formally ``completed".  Crucially, we make no assumption on the existence of a \textit{true} model, solely that the reference model passes posterior predictive checks. This stands in contrast to Bayesian model averaging for which one assumes an open model space \citep{vehtari2012}.}
Next, the posterior predictive information is projected optimally onto possibly parsimonious submodels. Previous implementations of projective predictive inference have not been considered for ARMA models and time-series models more generally, where the projection step is challenging due to the latency of the MA component and submodel search is complicated by a preference for exploring increasingly complex ARMA orders, however not at random. This is motivated by typical stationary time-series having rather smooth than ragged serial correlation structures, especially when any seasonal time-series correlations are appropriately modelled. To avoid selection of such ragged ARMA orders, we propose a forward search heuristic which iteratively increases the order until reaching the predictive performance of the reference model. This provides a safeguard against over-fitting, as the projected models typically won't exhibit better fit than that of the reference model.

The projection predictive paradigm differs from conventional selection approaches based on information criteria \citep{watanabe_widely_2013,spiegelhalter_bayesian_2002}%
% (\citep[WAIC;][]{watanabe_widely_2013}, \citep[DIC;][]{spiegelhalter_bayesian_2002})
, cross validation \citep{geisser_eddy_predictive_1979}, or subset selection from sparsity inducing priors \citep{barbieri_optimal_2004,horseshoe} in that selection is done with respect to point predictions from a reference model as opposed to the target data directly. \citet{piironen2016} and \citet{piironen2018} show that this improves stability of selection and is coherent with Bayesian decision theory rather than reliant on asymptotic approximations.\footnote{The WAIC is asymptotically equal to Bayesian LOO-CV, although both induce a small bias in their utility estimates. This can lead to high variance in the estimates of predictive utility, sub-optimal model selection, and over-fitting when the model space is large \citet{piironen2016}.} Although \citet{piironen2016} show that integrating over model uncertainty may provide the best predicting model, one may still use such a model in the proposed methodology as the reference for projection on more parsimonious submodels.

We aim to convince the reader of the benefits of using reference models for ARMA order selection by comparing the stability and predictive performance of reference model-based selection to that achieved by \pkg{auto.arima} \citep{autoarima}. We find through various simulation and real word data exercises that our procedure identifies models with predictive performance no worse than those found by \pkg{auto.arima}, and in some cases out-performs the latter. We also find that our procedure is robust to instances of noisy data, and avoids over-fitting in those situations where \pkg{auto.arima} liable to do so. By motivating a robust search heuristic, we show how our procedure is able to always produce well-performing models in cases where \pkg{auto.arima} fails.

This paper is organised as follows: we begin by discussing the theory behind ARMA models in Section~\ref{sec:tsa} before defining projection predictive model selection in Section~\ref{sec:projpred}. Having done so, we will be equipped to define our novel order identification procedure in Section~\ref{sec:projpred-arma} before discussing its position in the relevant literature in Section~\ref{sec:alternatives}. We then justify the utility of our procedure over the most widely used alternative through experiments in Section~\ref{sec:experiments} before summarising our contributions in Section~\ref{sec:conclusion}.

\section{Auto-regressive moving-average processes}\label{sec:tsa}
In this section, we briefly present the Bayesian ARMA model along with its decomposition, a central aspect of our proposed order selection approach later presented in section \ref{sec:projpred-arma}.
\subsection{The ARMA model}
Let the variable of interest $y_t$ be observed at integer time points $t \in \{1,\dotsc,T\}$ and assume its time series dynamics are described by an ARMA model of order $p, q$, denoted as $\arma(p, q)$:
\begin{equation} \label{eq:arma_gen}
    \phi(L)y_t = c + \theta(L)\varepsilon_t, \; \varepsilon_t \sim \normal(0,\sigma^2),
\end{equation}
where $\phi(L) = 1 - \phi_1L - \dotsc - \phi_pL^p$ and $\theta(L) = 1 + \theta_1L + \dotsc + \theta_qL^q$ are lag polynomials for the AR and MA component respectively, and $L$ is defined as the lag operator ($Lx_t = x_{t-1}$). We assume for simplicity throughout the paper that $c=0$ and that the initial conditions $\{y_0,\dotsc,y_{-p},\varepsilon_0,\dotsc,\varepsilon_{-q}\}$ are zero and given.\footnote{Note that one can treat these alternatively as unknowns to be estimated from the data if needed. However, when $T\gg \max(p,q)$, they have typically little impact on inference on the ARMA dynamics, even when ignored in the model.} Stationarity of Equation~\ref{eq:arma_gen} is not required to achieve proper posteriors via Bayesian updating with appropriate priors \citep{sims_bayesian_1988,sims_unitrooters_1991,schotman1991bayesian}.\footnote{These authors in fact show that the Bayesian approach is valid even when the AR and MA polynomials display unit roots, and are less prone to incur bias asymptotically for near unit-root data-generating processes.} In order to be able to identify the optimal ARMA orders for our projection algorithm, we will assume that the roots of $\phi(L)$ and $\theta(L)$ lie outside the unit circle and thus we have weak stationarity \citep{chib1994bayes}.\footnote{Note that the methods in this paper are amendable to non-stationary components, $\mu_t$ in the data-generating process, if one proceeds with $y^*_t = y_t-\mu_t$.}

To define the likelihood, we follow \citet{chan_moving_2013} and \citet{zhang2020stochastic} by first stacking all observations according to 
\begin{equation}
    H_{\phi}\boldsymbol{y} = H_{\theta}\boldsymbol{\varepsilon}, \; \boldsymbol{\varepsilon} \sim \normal(\boldsymbol{0},\Omega),
\end{equation}
where $\boldsymbol{\varepsilon} = (\varepsilon_1,\dotsc, \varepsilon_T)^T$, $\boldsymbol{y} = (y_1,\dotsc, y_T)^T$, $\Omega = \sigma^2 I_T$, and $H_{\phi}$ and $H_{\theta}$ appropriately defined as $T \times T$ difference matrices \citep{chan_moving_2013}. Since these are by definition lower triangular and invertible for any $\phi = (\phi_1,\dotsc, \phi_p)$, we can write $\boldsymbol{y} = H^{-1}_{\phi}H_{\theta}\boldsymbol{\varepsilon}$. Then the log-likelihood function is compactly written as
\begin{equation}\label{eq:log_lik_arma}
    \log(\boldsymbol{y}|\phi, \theta, \Omega) = -\frac{T}{2}\log(2\pi) - \frac{1}{2}|\Omega_y| - \frac{1}{2}(\boldsymbol{y}-\boldsymbol{c})^T\Omega_y(\boldsymbol{y}-\boldsymbol{c}),
\end{equation}
where $\Omega_y = H^{-1}_{\phi}H_{\theta}\Omega(H^{-1}_{\phi}H_{\theta})^T,\,\boldsymbol{c} \in\mathbb{R}^T$.

Importantly, the observational model of $\boldsymbol{y}$ belongs to the exponential family of distributions, namely the Gaussian, such that much of the theory for projection predictive inference in Section~\ref{sec:projpred} follows through immediately. The assumption of Gaussianity is made for simplicity and analytic tractability, although our proposed procedure makes no assumption on the observation family and thus could be naturally extended to other observational models, which we leave for future research.

To complete the model specification, we follow the recommendation of \citet{bayesforecast} who assume independent priors:
\begin{IEEEeqnarray*}{rl}
    \phi \;&\sim \normal(\boldsymbol{0},\Lambda_\Phi) \\
    \theta \;&\sim \normal(\boldsymbol{0},\Lambda_\Theta) \\
    c \;&\sim \student t(0,\sigma_c,\nu) \\
    \sigma \;&\sim \student t_+(0,\sigma_{\sigma},\nu_{\sigma}), \label{eq:gaussian-arma}
\end{IEEEeqnarray*}
where $\student t_+(\cdot,\cdot,\cdot)$ refers to the half-Student-$t$ distribution with positive support, and $\Lambda_\Phi,\,\Lambda_\Theta$ denote the diagonal prior covariances motivated by \citet{bayesforecast}. While a plethora of variable selection and shrinkage priors have been proposed for highly parameterised time series forecasting models such as vector auto-regressive (VAR) models \citep{banbura2010large,koop2013forecasting,carriero2015bayesian,giannone_prior_2015,chan2016large}, this is less of a concern for relatively more parsimonious ARMA models. In this paper, we purposefully consider a statistician who uses default priors to compare our procedure with alternative (potentially non-Bayesian) ARMA selection techniques such as the popular \pkg{auto.arima}. 

Independently to the choice of priors, it is well known that ARMA likelihoods can be multi-modal \citep{Chan2011SubsetAS} and additionally pose computational challenges due to the latent MA components.\footnote{Traditionally used Kalman filters for maximum likelihood formulations of ARMA models \citep{harvey1985trends} can become computationally demanding, particularly with large $T$  \citep{kim1999state}.} To aid posterior inference, all models in this paper are estimated via Hamiltonian Monte Carlo \citep{neal2011mcmc} using the so-called no U-turn sampling (NUTS) algorithm as implemented in \proglang{Stan} \citep[version 2.26.1;][]{hoffman_no-u-turn_2011,carpenter2017stan}. 
\subsection{Decomposing the ARMA model}\label{sec:arma-linear}
ARMA models pose the computational problem that the MA component $\theta(L)\varepsilon_t$ are unobserved. To deal with this, we follow the logic presented in least-square literature on ARMA estimation \citep{kapetanios2003note} by splitting the $\arma(p,q)$ model into its auto-regressive and moving average components and then estimating these in two separate steps as proposed by \citet{hannan1982recursive}. 

To illustrate this, assume oracle knowledge of the order from the respective 
AR and MA components, and begin by fitting $\phi(L)y_t = \delta_t$, where $\delta_t$ is some white noise process, without observing $\varepsilon_t$. Then, we fit a linear model to the residuals, $\hat{\delta}_t$, to approximate the MA component. Formally,
\begin{IEEEeqnarray}{rl}
    \phi(L)y_t \;&= \delta_t \nonumber\\
    \theta(L)\hat{\delta}_t \;&= \xi_t,\nonumber
\end{IEEEeqnarray}
noting that $\hat\delta_t\approx\varepsilon_t$ from Equation~\ref{eq:arma_gen}, and $\xi_t$ are from a white noise. \citet{ng1995unit} detail the sufficient conditions required to estimate $\hat\delta_t$ consistently, which in turn allows for consistent estimation of the ARMA order from a frequentist perspective. In performing this sequential model fitting, we also simplify our search heuristic when traversing the model space as we shall see later in Section~\ref{sec:search}, and forms the basis of our two-step procedure later defined in Section~\ref{sec:projpred-arma}.
\subsection{Seasonal ARMA models}
One innovation on the base ARMA model previously discussed is to model recurring seasonal trends. We can do so by defining structural relationships occurring at regular lag intervals, for example each week, quarter, year, and so on. The multiplicative seasonal auto-regressive moving-average (SARMA) model with seasonal patter repeating every $s$ lags, non-seasonal ARMA components $p, q$ and seasonal components $P, Q$ is denoted $\arma(p, q)\times(P, Q)_s$ and written in terms of polynomials in $L$, as
$$
\Phi(L^s)\phi(L)y_t = \Theta(L^s)\theta(L)\varepsilon_t.
$$
These SARMA models extend the theory of ARMA models to fit and forecast more complex, macro trends in time series data and thus lend themselves to more valuable application.
\section{Projection predictive model selection}\label{sec:projpred}
In this section we outline the underlying theory of projection predictive inference, as well as our contributions to this approach in making it amenable to ARMA models. 
\subsection{The projection}\label{sec:projpred-idea}
The idea of projection predictive inference is to separate prediction and model selection into two stages: we first identify a model that produces the best predictive performance given the information set of the statistician, which we call the \textit{reference model} \citep{vehtari2012}; we then construct smaller models capable of replicating the predictive performance of this reference model. These smaller models are essentially fit to the fit of the reference model (a procedure we call the \textit{projection}), and are then used for their improved interpretability, or to decrease data collection cost \citep{piironen2018}. We usually identify these smaller models via sparsity-favouring search heuristics \citep{hahn2015decoupling,ray2018signal,kohns2022flexible}. Using a reference model in the model selection process has been shown previously by \citet{piironen2016}, and more recently by \citet{Pavone2020}, to lead to more stable submodel discovery in which submodels are less prone to over-fit the data than those found through other procedures that conduct selection directly on $y_t$. A salient reason for the stability of the projection is the fact that a well-defined reference model is able to separate signals of the data-generating process from noise \citep{piironen2018}.

We start, then, with the model containing all available data which we fit using reasonable priors, and which has passed posterior checks \citep{gelman1996posterior,bayesianworkflow,gabry_visualization_2019}. This model is written in terms of the full parameter space $\boldsymbol\theta^\ast\in\boldsymbol\Theta^\ast$, and is fit to the observed data $\mathcal{D} = \{X, y\}$. From this reference model, we wish to achieve some more parsimonious model in the restricted parameter space $\boldsymbol\theta^\perp\in\boldsymbol\Theta^\perp$ that will often incorporate sparsity.\footnote{In general, we do not require this restricted parameter space to be a subset of the reference parameter space, although it is usually chosen to induce sparsity in the submodels.} Concretely, we replace the posterior distribution over the reference model parameters $p(\boldsymbol\theta^\ast\mid\mathcal{D})$ with some simpler distribution $q(\boldsymbol\theta^\perp)$ such that its induced covariate-conditional predictive distribution, $q(\tilde{y}\mid\boldsymbol\theta^\perp)$, is not significantly different to that of the reference model. We quantify this with some distance measure,
\begin{equation}
    d\left\{p(\tilde{y}\mid\mathcal{D}, \boldsymbol\theta^\ast),\, q(\tilde{y}\mid\boldsymbol\theta^\perp)\right\} \leq \varepsilon, \label{eq:predictive-dist}
\end{equation}
where $\tilde{y}$ represents predictions of the variate, $d$ is a distance measure, and $\varepsilon>0$ is small. \citet{vehtari2012} consider projected posteriors from a decision-theoretic standpoint and reason that given a logarithmic utility function, the optimal values of the restricted model parameters $\boldsymbol\theta^\perp$ are achieved by minimising the Kullback-Leibler (KL) distance between the posterior predictive distributions of the reference and restricted model. This divergence benefits from its analytical tractability and efficient computation within the exponential family of distributions, as well as the guarantee of a unique optimum and natural position within a Bayesian workflow \citep{goutis1998, piironen2018}.

Assume we have collected $S$ posterior draws from the reference model, $\{\boldsymbol\theta_{(s)}^*\}_{s = 1}^S$. 
The projection is then simply the solution to the optimisation problem,
$$
\boldsymbol\theta^\perp_{(s)} = \argmin_{\boldsymbol\theta\in\boldsymbol\Theta^\perp}\mathbb{KL}\left\{\, p(\tilde{y}\mid\mathcal{D}, \boldsymbol\theta^\ast_{(s)}) \,\vert\vert \,q(\tilde{y}\mid\boldsymbol\theta_{(s)}) \,\right\},
$$
for which one can show analytical solutions exit when the models are contained within the exponential family of distributions. \citet{piironen2018} propose to solve this problem sample-wise for computational reasons, leaving us with some set of projected samples we can consider as samples from the submodel's posterior (and which we achieved at a negligible cost since no Monte Carlo methods were used in projection step). We may then evaluate the new submodel's predictive performance with such metrics as expected log-predictive density \citep[elpd;][]{loo} using leave-one-out cross-validation (LOO-CV) to compare it to the reference model.

Since LOO-CV is computationally demanding, particularly when $T$ is large, we will use the recently proposed approximate LOO-CV using Pareto smoothed importance sampling \citep[PSIS;][]{loo}. This represents a fully probabilistic way of doing LOO-CV that avoids repeatedly fitting the reference model by re-weighting the posterior draws $\{\boldsymbol\theta_{(s)}^*\}_{s = 1}^S$ with importance weights. In particular, the weight for draw $\boldsymbol{\theta}^*_{(s)}$, leaving the $t^{th}$ out, denoted $\omega_{(s)}^{-t}$ where $t$ indexes the left-out observation, is given by 
\begin{equation}
    \omega_{(s)}^{-t} \propto \frac{1}{p(y_t|\boldsymbol{\theta}^*_{(s)})}. \nonumber
\end{equation}
These weights are then stabilised with Pareto smoothing for instances in which the importance weight distribution has a thick tail \citep{loo}.
\subsection{Search strategies for traversing the ARMA model space}\label{sec:search}
\begin{figure*}[t!]
    \begin{subfigure}{\textwidth}
        \centering
        \resizebox{0.8\textwidth}{!}{%
            \begin{tikzpicture}[grow'=right,level distance=1.25in,sibling distance=.25in]
                \Tree 
                    [. {$\small \{\text{Intercept}\}$ }
                        [.{\small $\{\text{Intercept}, \phi_{1}\}$}
                            [.{\small $\cdots$ } ]
                            [.{\small $\cdots$ } ]
                            [.{\small $\{\text{Intercept}, \phi_{1}, \phi_{4}\}$ } 
                                [.{\small $\cdots$ } ]
                                [.{\small $\cdots$ } ]
                                [.{\small $\cdots$ } ]
                            ]
                        ]
                        [.{\small $\cdots$ } ]
                        [.{\small $\cdots$ } ]
                    ]
            \end{tikzpicture}
        }
        \caption{A forward search heuristic in the auto-regressive space, in which each step considers all previously unconsidered parameters to be appended to the path regardless of their lag or the feasibility of the resulting model.}
        \label{fig:forward-search}
    \end{subfigure}
    
    \vspace{2em}
    \begin{subfigure}{\textwidth}
        \centering
        \resizebox{0.8\textwidth}{!}{%
            \begin{tikzpicture}[grow'=right,level distance=1.25in,sibling distance=1.25in]
                \Tree 
                    [. {\large $\{\text{Intercept}\}$}
                        [.{\large $\{\text{Intercept}, \phi_{1}\}$}
                            [.{\large $\cdots$ } 
                                [.{\large $\{\text{Intercept}, \phi_{1}, \phi_{2}, \phi_{3}, \phi_{4}\}$ } 
                                [.{\large $\cdots$ } ]
                            ]
                            ]
                        ]
                    ]
            \end{tikzpicture}
        }
        \caption{The proposed search path through the parameter space, where each next parameter is that one situated just one lag further from the observed point, thus considering only valid ARMA models.}
        \label{fig:arma-search}
    \end{subfigure}
    \caption{A comparison of parameter search heuristics discussed in Section~\ref{sec:search}. Each node in the tree represents a submodel's parameters, and the path from the intercept-only model to the final model (from left to right in the graphs) represents the search path.}
\end{figure*}
In order to determine which parameter subsets to project our reference model onto, we need some search heuristic to propose a collection of submodels. \citet{piironen2018} propose a forward search heuristic in the case of generalised linear models, wherein candidate submodels are found by iteratively appending the variable minimising the distance in Equation~\ref{eq:predictive-dist} starting with the intercept-only model. Such search heuristics do not adapt well to the ARMA model. 

For instance, suppose we have an $\ar(1)$ model and  to it we add the $\phi_{4}$ parameter in our search. In doing so, one may be prone to skipping intermediate lags which creates a seasonal behaviour in the AR dynamics. This, however, should not appear, particularly when modelling seasonal components as we do in section \ref{sec:sarma} and making sure that outliers and departures from the stationary component of the data-generating process are adequately modelled in the reference model. This example is seen in Figure~\ref{fig:forward-search}.

Different from order identification schemes such as \citet{autoarima}, we implement a search heuristic that iteratively appends the next lagged variable to the set, since this is directly equivalent to increasing the order of our AR or MA model by one with each added parameter. For example in Figure~\ref{fig:arma-search}, the submodel containing $\phi_{4}$ is indeed an $\ar(4)$ model. 

Alternative order selection procedured proposed by \citet{nardi_autoregressive_2011} and \citet{Chan2011SubsetAS} instead conduct variable selection directly on the ARMA polynomials via frequentist lasso style regularisation. They prove that under some regularity conditions, an adaptive Lasso regression of the time series on its lags enjoys oracle properties asymptotically. We do not consider this any further since we remain primarily interested in the finite data regime.
\subsection{Submodel acceptance heuristics}\label{sec:select}
Having projected our reference model onto different parameter subsets, we then move on to identify the smallest submodel whose predictive performance is comparable to that of the reference model. We propose the submodel selection heuristic outlined by \citet{piironen2016} based on the sensible cross-validation of submodels following a forward search through the parameter space. Denote the elpd of the reference model as $\elpd(\mathcal{M}^\ast\mid y)$ and that of the submodel on $p$ parameters as $\elpd(\mathcal{M}_p\mid y)$. Having projected our reference model onto a set of submodels, we choose the submodel with the smallest $p$ for which the upper bound of its normal-approximation $68\%$ elpd confidence interval (the one standard deviation range) is at least as good as the elpd point estimate of the reference model:
\begin{equation}
    \mathbb{P}\left(\elpd(\mathcal{M}^\ast\mid y) - \elpd(\mathcal{M}_p\mid y) \leq 0\right) \geq 0.68. \label{eq:selection}
\end{equation}
\section{Projection predictive ARMA order identification}\label{sec:projpred-arma}
Having now addressed both projection predictive model selection and the ARMA model, we combine the former with the decomposition of ARMA models outlined in Section~\ref{sec:arma-linear}, the selection heuristic proposed in Equation~\ref{eq:selection} and the search heuristic from Section~\ref{sec:search} to build our order identification procedure.

\subsection{A fully Bayesian ARMA order identification procedure}\label{sec:arma-procedure}
Suppose we have some data we wish to model with an ARMA model
\begin{equation*} 
    \phi(L)y_t = c + \theta(L)\varepsilon_t,
\end{equation*}
where $\{\varepsilon_t\}$ are some noise in our data and assume that we have fit an ARMA reference model that passes predictive checks.\footnote{More generally, one might consult the empirical auto-correlation functions to determine the maximum $\arma(p,q)$ to start the analysis.}. Denote the orders of the reference model as $(p^*,q^*)$. Taking $(\{ y_t \}, p^\ast, q^\ast)$ as inputs, we propose the application of projection predictive inference as outlined in Algorithm~\ref{algo}, and which we briefly describe below.\footnote{An implementation of Algorithm~\ref{algo} in \proglang{R} is available at \url{https://github.com/yannmclatchie/projpred-arma}.}

\begin{algorithm}[t!]
 \setstretch{1.15}
 \caption{Projection predictive order identification of ARMA models}\label{algo}
 \begin{algorithmic}[1]
 \Procedure{ProjpredARMA}{$\{ y_t \}, p^\ast, q^\ast$}
 \vspace{2ex}
 \State Fit an auto-regressive model with HMC to data observations $\{ y_t \}$ with the lag $p^\ast$ from the reference model, $\ar(p^\ast)$, using MCMC and reasonable priors.
 \vspace{2ex}
 \State Treat this $\ar(p^\ast)$ model as the auto-regressive reference model, and apply projection predictive model selection to it using the temporal search method defined in Section~\ref{sec:search} and extract the restricted lag value $p^\perp$ with the selection heuristic defined in Equation~\ref{eq:selection}.\label{step:projpred}
 \vspace{2ex}
 \State Fit a linear model to the residuals of the $\ar(p^\perp)$ model $\{ \varepsilon_t^\perp \}$ (through posterior predictive mean point estimates) with HMC, and using the reference model MA lag, $q^\ast$. \label{step:ma-fitting}
 \vspace{2ex}
 \State Apply the same modified projection predictive model selection as in step~\ref{step:projpred} to this linear reference model of order $q^\ast$ from step~\ref{step:ma-fitting} and retrieve $q^\perp$.
 \vspace{2ex}
 \State \textbf{return} $p^\perp, q^\perp$.
 \vspace{2ex}
 \EndProcedure
 \end{algorithmic}
\end{algorithm}

First, refit an $\ar(p^*)$ model to $y_t$. For instance, should we find that a reference $\arma(5, 5)$ fits the data well, then we would fit a AR reference model of structure $\ar(5)$.

Second, we perform projection predictive model inference on this $\ar(5)$ model to find a possibly more parsimonious restricted model of order $\ar(p^{\perp})$, say an $\ar(1)$ for illustration, using the temporal search heuristic in \ref{eq:selection}. We save the residuals of the projected AR model, denoted as $\{ \varepsilon_t^\perp \}$ since these will serve as the observed errors to project the MA component next. Note that this step produces an implied distribution over the residuals upon prediction. For simplicity, we define $\{ \varepsilon_t^\perp \}$ here as the posterior mean of the residuals.

Third, we fit an linear model of the order defined by the reference model (in this case $q^\ast = 5$) to the residuals $\{ \varepsilon_t^\perp \}$.  

Finally, we perform projection predictive model inference on this linear model fit to the residuals to retrieve a restricted MA order $q^{\perp}$ (reducing an $\ma(5)$ to, say, an $\ma(3)$). The combined orders of the restricted submodels identify the order of the projected $\arma(p^{\perp},q^{\perp})$ that most closely replicates the predictive performance of the much larger $\arma(p,q)$ reference model as measured by Kullback-Leibler distance.

This methodology allows us to search the parameter space in an intuitive manner in terms of two linear models sequentially and with minimal information loss while benefiting fully from the stability afforded by a reference model.

While these $p^\ast$ and $q^\ast$ can in theory take any value, we will limit them in our experiments to $p^\ast = q^\ast = 5$ in line with the default values proposed by \citet{autoarima}. Indeed, if we were to set these reference lags to larger values, we would have to choose our priors appropriately to communicate the fact that more distant lags are less likely to have an effect on the present and in order to enforce model stationarity.
\subsection{Extension to SARMA models}\label{sec:sarma}
We presently show how we might naturally extend the procedure presented in Algorithm~\ref{algo} to SARMA models. In this scenario, we define our reference model as a function of the seasonality being modelled. Specifically, given a reference model $\arma(p^\ast, q^\ast)\times(P^\ast, Q^\ast)_s$, with both seasonal and non-seasonal components and seasonality $s$, we produce two datasets: one seasonal and one non-seasonal to which we apply our procedure independently. The outputs of these two runs provide us the non-seasonal and seasonal restricted parameters respectively. Again, in line with \citet{autoarima}, we choose as default values $P^\ast = Q^\ast = 3$ in our experiments.
\subsection{Cross-validation with time series data}
In previously seen experiments with projection predictive model selection, such as those carried out by \citet{piironen2016}, \citet{piironen2018}, \citet{catalina2021latent}, and \citet{projpredgam}, evaluation of each of the submodels was performed with LOO-CV for efficiency. Approximate leave-future-out cross-validation \citep[LFO-CV;][]{lfo} provides an alternative to this for time series data cross-validation. 

In order to convince the reader that the computationally cheaper LOO-CV is justified in our case, consider a time series model $p(y_i \mid f_i, \boldsymbol\theta)$ with latent process values $\boldsymbol f$ and prior $p(\boldsymbol f)$ such that $(y_i - f_i)$ are independent and identically distributed. Now, if we are interested in predicting unseen future observations, then LFO-CV is the natural choice. However, should we instead be interested in reasoning on the structure of our time series, that is the conditional observation model $p(y_i \mid f_i)$, then \citet{lfo} show that it is reasonable to use LOO-CV instead. The authors reason that LOO-CV can be thought of as a biased approximation to LFO-CV, and further that the biases in LOO-CV and LFO-CV are likely in the same direction \citep{lfo}. This means that since the difference between model comparison through LOO-CV and LFO-CV is small, and since it has been empirically noted that covariate ordering is similar between the two, we are able to rely on the computationally cheaper LOO-CV for variable selection in time series models.

The use of the AIC and the Bayesian information criterion \citep[BIC;][]{schwarz_estimating_1978}, both commonly implemented in analyses, are themselves asymptotically equivalent to LOO-CV and $K$-fold-CV respectively \citep{stone1977asymptotic,shao1997asymptotic,arlot2010survey,vehtari2012}. As such, using LOO-CV can be considered the finite-sample analogue of the asymptotically equivalent AIC.
\subsection{Alternative projections}\label{sec:arma-to-ar-proj}
In general, projection predictive inference accommodates the possibility of projecting an arbitrary reference model structure onto a set of models with different structures. We briefly discuss how our procedure in Algorithm~\ref{algo} could be adapted to achieve differently nuanced model selection results.

We might be tempted, for example, to project our reference ARMA model directly onto the AR component rather than perform the initial decomposition, and then continue the procedure from step~\ref{step:ma-fitting} of Algorithm~\ref{algo}. We would expect this to then over-select the AR size and under-select the restricted MA component, since we then project the information communicated by both the AR and MA onto solely the AR component of the restricted model. There do exist, however, instances where such projections may afford the statistician a reduced computational cost. \citet{chan_moving_2013} discusses the difficulty of inferring MA parameters often present in economic models. As a remedy to this, and understanding that the inclusion of an MA component in the ARMA model induces an infinite AR order, one might project their ARMA model onto only an AR component purposefully to achieve the minimal order necessary to replicate the behaviour of an $\ar(\infty)$ model at hopefully less computational cost.

Further, if instead of finding parsimonious submodels, our interest is in identifying the true order of the ARMA with our approach, one might consider application of so-called complete variable selection as presented by \citet{Pavone2020}, where the statistician is interested in identifying all covariates (theoretically) relevant to predicting the target. Here, the projective inference step is repeatedly applied until some stoppage criterion is met.\footnote{\cite{Pavone2020} suggest using local false discovery rates \citep{efron_microarrays_2008,efron_large-scale_2010}, empirical Bayes median \citep{johnstone_needles_2004}, and posterior predictive credible intervals.} Such an approach can similarly be applied to update the projected residuals with each projection step.

We show how these slightly different projection techniques can be implemented in Section~\ref{sec:bad-proj}. We remain primarily interested in identifying parsimonious ARMA orders in this paper, and so do not consider complete variable selection any further.
\section{Related work}\label{sec:alternatives}
Other model selection procedures exist in the case of ARMA models. We brielfy discuss the most competitive alternative to our procedure in \pkg{auto.arima}, and motivate why other procedures are inadequate to identify and communicate ARMA subset selection.

\subsection{Automatic order selection with unit root tests and AIC}\label{sec:autoarima}

The \pkg{forecast} package in \proglang{R}, developed by \citet{autoarima}, has long been the \textit{modus operandi} of statistical practitioners for the order identification of ARMA models. In particular, the \pkg{auto.arima} module automates an order selection heuristic based on unit root tests and the AIC.

In the original algorithm proposed by \citet{autoarima}, some unit tests are performed on the data and based on their results, they fit four models by maximum likelihood. These four initial models are pre-defined before we see any data and without any prior information. Of these four models, they select the one with the lowest (best) AIC, and consider a further thirteen variations of it. The model with the lowest AIC of these resulting models is chosen as the best submodel. Various constraints are imposed to ensure convergence or near unit root, and these constraints combined with the finite search space guarantees that at least one of the models considered will be valid \citep{autoarima}. 

Nevertheless, due to its competitive performance and popularity, we will consider \pkg{auto.arima} for comparison in later experiments. As such, we use the default algorithm in \citet{autoarima} to identify the ARMA orders and then conduct inference via Hamiltonian Monte Carlo, using the same priors as in Algorithm \ref{algo} so that we can compare predictive performance with our procedure. This is summarised in Algorithm~\ref{algo:autoarima}.

Contrasting Algorithms~\ref{algo} and~\ref{algo:autoarima}, we highlight two main reasons why projective inference may result in most stable submodel discovery. Firstly, it is well known that priors can have a regularising impact on the posterior, and that MCMC methods such as HMC allow us to explore the posterior more efficiently than maximising the likelihood directly. Secondly, projective inference conducts submodel selection based on predictions of a reference model, whereas \pkg{auto.arima} conducts its selection directly on the observations $\{y_t\}$. This use of a reference model helps submodel discovery by typically filtering out noise from the observations, and can also help avoid over-fitting to the data since the fit of the submodels is bounded by that of the reference model.\footnote{In fact \citet{mcquarrie} note the tendency of AIC-based model selection to over-fit data in instances of small sample size.} This will be formally investigated in Section~\ref{sec:stability}.
\begin{algorithm}[t!]
 \setstretch{1.15}
 \caption{MCMC fitting of an \pkg{auto.arima}-selected model.}\label{algo:autoarima}
 \begin{algorithmic}[1]
 \Procedure{MCMC AutoARIMA}{$\{ y_t \}$}
 \vspace{2ex}
 \State Identify some orders $p_{\auto}, q_{\auto}$ from differenced, stationary data with \pkg{auto.arima}.
 \vspace{2ex}
 \State Fit the $\arma(p_{\auto}, q_{\auto})$ model with previously found orders using MCMC, employing some reasonable priors.
 \vspace{2ex}
 \EndProcedure
 \end{algorithmic}
\end{algorithm}
\subsection{Cross-validation}
Another common approach to model selection is to fit some collection of models and estimate their respective elpd scores \citep[or any other scoring rule, e.g. those proposed by][]{gneiting2007strictly} with cross-validation, whereupon the optimal subset according to the highest score is chosen.

While this procedure has gained popularity \citep{arlot2010survey}, and while it has been shown to be a robust method when dealing with relatively few models, if the number of models being compared is relatively large or the number of covariates is relatively small, various issues arise. \citet{piironen2018} showed that when the number of models is large, then cross-validation without a reference model is liable to over-fit and result in the selection of a sub-optimal model. Indeed, \citet{piironen2016} compared this approach directly with projection predictive model selection, finding that the latter is significantly more resilient to these issues. \citet{Pavone2020} further showed that the use of a reference model in model selection affords a greatly improved stability in selection which is due to that fact that the reference model is able to filter out noise before arriving at the submodel selection stage.

A computational hurdle for Bayesian workflows is additionally that full cross-validation requires fitting many models for which MCMC is re-conducted for each left-out observation. This is a highly expensive endeavour. To summarise, we present in Table~\ref{tab:complex} algorithmic complexities of the reviewed methods.

\begin{table}[t!]
    \centering
    \begin{tabular}{lc}\toprule
    \textbf{Procedure}  & \textbf{Worst-case complexity} \\ \midrule
    ProjpredARMA & $\mathcal{O}(1)$ \\
    MCMC AutoARIMA & $\mathcal{O}(1)$ \\
    Cross-validation & $\mathcal{O}(p^\ast \cdot q^\ast)$ \\ \bottomrule
    \end{tabular}
    \caption{A comparison of the algorithmic complexity of the four proposed model selection procedures as a function of the number of models needed to be fit by MCMC.}
    \label{tab:complex}
\end{table}
Since fitting a model with MCMC represents the largest user and computational cost, we measure the complexity of each procedure as a function of the number of models needing to be fit by MCMC and present their worst case algorithmic complexities.
In ProjpredARMA the number of models needed to be fit by MCMC is independent of the size of the reference model, namely, we will always fit two models (AR and MA components) with MCMC regardless of the maximum lags $p^\ast$ and $q^\ast$. Similarly, in our MCMC AutoARIMA procedure, we leverage the speed of \pkg{auto.arima} to only fit one model by MCMC. However, when using a pairwise selection criterion such as cross-validated elpd, in the worst case we need to fit all possible models encompassed by maximum lags $p^\ast$ and $q^\ast$. Consequently we find that the number of models needed to be fit has complexity $\mathcal{O}(p^\ast \cdot q^\ast)$.
\subsection{Using sparsifiying priors for ARMA models}\label{sec:sparse}
A commonly used alternative to variable selection for generalised linear models are sparsifying priors over the regression coefficients such as the regularised horseshoe prior \citep{horseshoe}, the spike-and-slab prior \citep{spikeslab} or R2D2 prior \citep{r2d2} \citep[see e.g.][for excellent reviews on further shrinkage priors]{polson2012half,bhadra2019lasso}. Such priors force certain parameter values close to zero based on their relevance to posterior predictions. 

Such priors face issues in our case. Namely, as is noted by \citet{catalina2021latent}, the posterior of a model fitted with a sparsifying prior is not truly sparse in that parameter posteriors are not generally point masses at zero, and manual effort is required from the statistician to first conceive a threshold of posterior relevancy, and then to prune those parameters beneath it. Then, it is not clear that these sparsified posteriors represent intuitive or desirable ARMA models in general, as such priors are usually used under the assumption of exchangeability of the regression weights. When the data are highly correlated, as we expect with AR and MA lags, aggressive shrinkage priors may cause the marginal posteriors of the regression weights to overlap strongly with zero, thus falsely indicating insignificant lags. For instance, we might identify that the first three lags may be pruned based on the individual marginal posteriors, yet they are important to include so as not to create unintended seasonal patterns in the predicted ARMA (much like in Figure~\ref{fig:forward-search}). And in the particular case of the spike-and-slab prior, the optimality of selecting the median probability model for prediction assumes orthogonality of covariates, which is not the case in ARMA models. Indeed \citet{barbieri_optimal_2004} put forward a case in which the median probability model with correlated covariates is clearly sub-optimal. We therefore do to not consider sparsifying priors any further as an alternative model selection procedure in our case. For a comprehensive comparison of projection predictive inference with sparsifying priors and median probability model, and a comparison with projection predictive inference, see the review by \citet{piironen2016}.

Alternatively to using shrinkage priors, the so-called Minnesota prior \citep{giannone_prior_2015} and its adaptive variant \citep{chan2021minnesota} can be used to shrink lag effects as a function of their distance from the current realisation through time. One might also consider functional restrictions on AR and MA lag polynomials such as used in mixed-frequency applications \citep{kohns2022flexible,mogliani2021bayesian}. While useful to some applications, such restrictions create irreducible bias when the restrictions are not approximately correct. If, however, sufficient knowledge of such restrictions exist, they may provide an adequate reference model within the proposed framework.
\section{Experiments}\label{sec:experiments}
We presently demonstrate the practical value of our proposed procedure compared to \pkg{auto.arima}. We do so first by using it to identify predictive submodels from multiple simulations of different data-generating processes and comparing the stability of submodel selection of the two procedures, as well as their closeness to the true process. We then fit some reference models to a selection of well-known datasets and compare our procedure to \pkg{auto.arima} in achieving predictive submodels. Having investigated the procedures' stability and the predictive performance of their submodels, we then compare the performance of the two procedures under different noise regimes. Finally, we illustrate the behaviour of the search heuristic motivated in Section~\ref{sec:select} in an auto-regressive example with many distant, near-zero lags compared to \pkg{auto.arima}, and conclude with a demonstration of how our procedure scales to larger data.

The models used in the experiments were fitted with the defaults priors suggested by \citet{bayesforecast} and all experiments were performed with a modified version of \pkg{projpred} \citep{projpredpackage}.
\subsection{Stability in model selection} \label{sec:stability}
\begin{figure*}[t!]
\centering
\begin{subfigure}{0.475\textwidth}
    \includegraphics{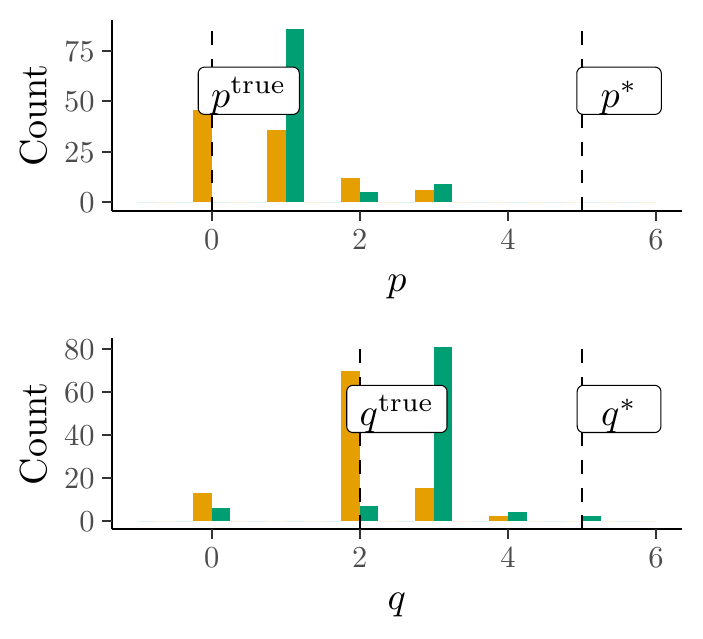}
    \vspace{-1em}
    \caption{$\text{ARMA(0,2)}$ model.}
    \label{fig:arma(0,2)}
\end{subfigure}
\hfill
\begin{subfigure}{0.475\textwidth}
    \includegraphics{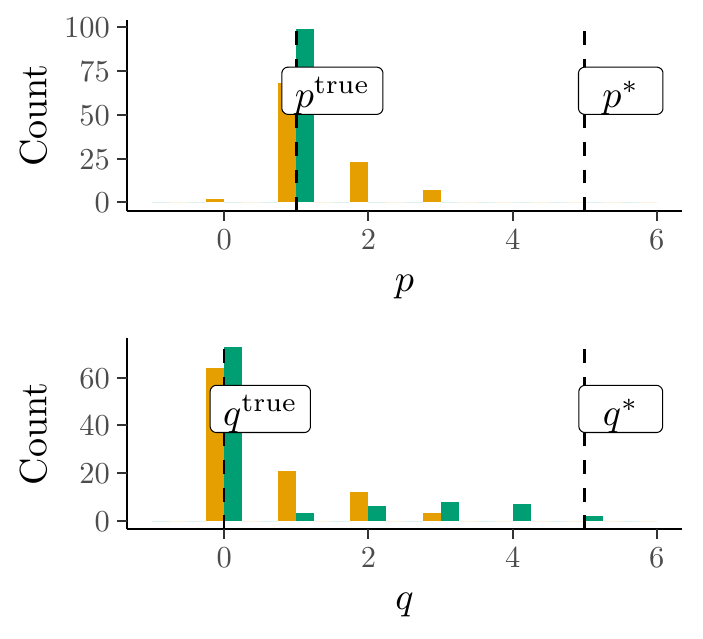}
    \vspace{-1em}
    \caption{$\text{ARMA(1,0)}$ model.}
    \label{fig:arma(1,0)}
\end{subfigure}
\\
\vspace{5ex}
\begin{subfigure}{0.475\textwidth}
    \includegraphics{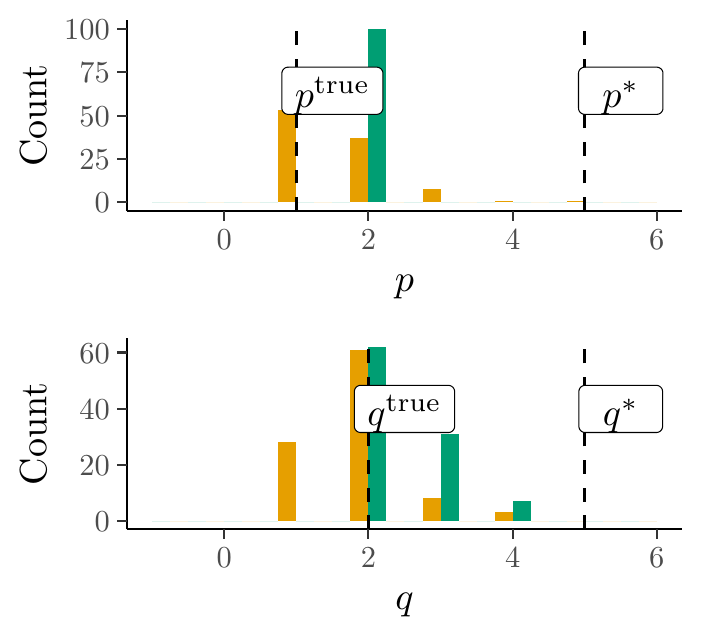}
    \vspace{-1em}
    \caption{$\text{ARMA(1,2)}$ model.}
    \label{fig:arma(1,2)}
\end{subfigure}
\hfill
\begin{subfigure}{0.475\textwidth}
    \includegraphics{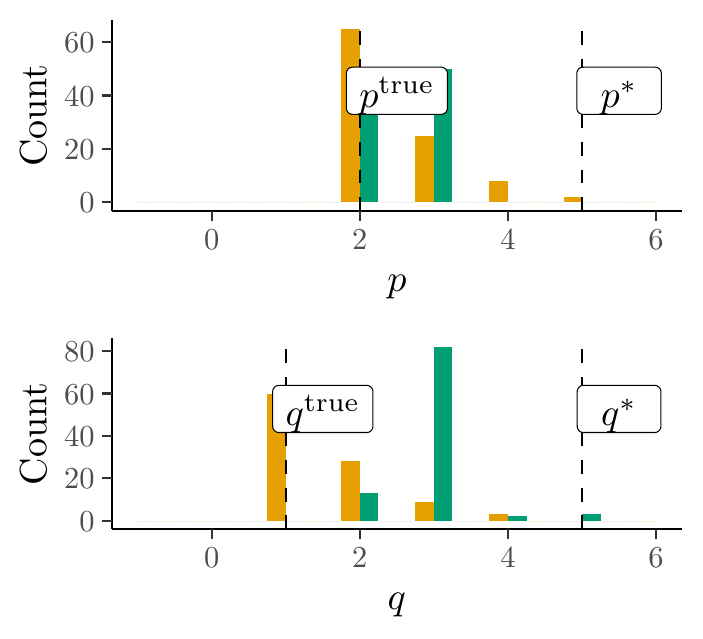}
    \vspace{-1em}
    \caption{$\text{ARMA(2,1)}$ model.}
    \label{fig:arma(2,1)}
\end{subfigure}
\caption{Simulated data. Histograms showing the frequency of model orders selected by projection predictive inference ($p^{\perp}$ and $q^{\perp}$ in \textcolor{projpredcolour}{green}) compared to \pkg{auto.arima} ($p_{\auto}$ and $q_{\auto}$ in \textcolor{autocolour}{orange}). The reference model parameters used, $p^\ast$ and $q^\ast$, and the true data-generating parameters $p^\text{true}$ and $q^\text{true}$ are labelled and shown by the dotted black lines. We find that across non-seasonal simulated data experiments, projection predictive inference benefits from far greater stability than \pkg{auto.arima}.}
\label{fig:arma-results}
\end{figure*}
\begin{figure*}[t!]
    \centering
    \includegraphics{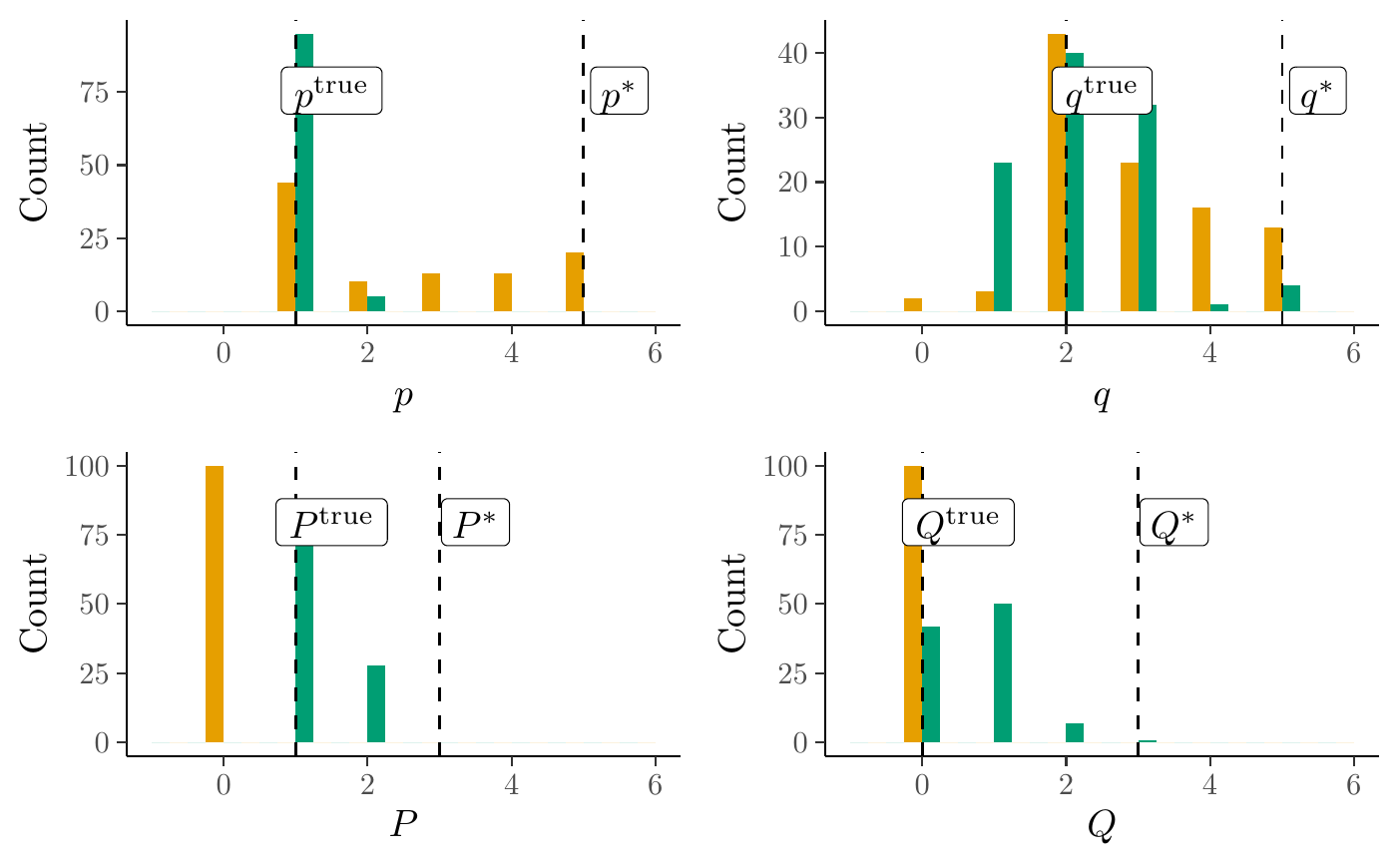}
    \caption{Simulated data. Histograms showing the frequency of model orders selected by projection predictive inference (in \textcolor{projpredcolour}{green}) compared to \pkg{auto.arima} (in \textcolor{autocolour}{orange}) for data sampled from a $\text{SARMA}(1,2)\times(1,0)_{12}$ model. The reference model parameters and the true data-generating parameters are shown by the dotted black lines. We find that projection predictive inference again benefits from increased stability compared to \pkg{auto.arima} across seasonal simulated data experiments.}
    \label{fig:sarma-results}
\end{figure*}
We presently simulate data according to four $\arma(p, q)$ processes, each with different orders $p$ and $q$, and aim to recover some restricted model with predictive performance close to a reference $\arma(5, 5)$ model. Namely, the true data generating processes we sample from are the $\arma(1, 0),$ $\arma(0, 2)$, $\arma(2, 1)$, and $\arma(1, 2)$ models. We will then demonstrate the validity of the extension of the procedure to SARMA processes by repeating this for an $\arma(1,2)\times(1,0)_{12}$ model. For each model, we run $100$ simulations each generating $500$ data points. For each of these simulated series, we employ both projection predictive inference and \pkg{auto.arima} to identify parsimonious model orders. 

We define stability in model selection concretely in terms of the concentration of the distribution of model orders selected. In a word, we call a model selection procedure ``stable'' if the orders of the ARMA model selected are similar across repetitions.

In Figure~\ref{fig:arma-results} we see that the auto-regressive orders identified by projection predictive inference are significantly more stable than those selected by \pkg{auto.arima}. The moving-average component is also broadly more stable, but not as significantly so. This is likely a symptom of the model decomposition we propose, where some noise is perhaps leaking into the residuals of the AR component and contaminating our decision. We further note that these highly-stable values of $p^\perp$ and $q^\perp$ are either precisely the true model values or were close to the true values. When our procedure ``incorrectly" selects lag values (by which we mean not exactly the true model), most selections still fall within one lag from the actual value, showing that our procedure has learned some of the underlying structure in the time series data even in the worst cases. The AR component retrieves exactly the correctly lag in the absence of an MA component in the ARMA data-generating process. Indeed, since the inclusion of any MA component leads to an infinite AR component, it is expected that our restricted AR order will marginally over-select -- as is the case. We include this analysis to demonstrate that despite the aim of the procedure not explicitly being about retrieving a true model (since we do not assume one exists), it is able to identify some underlying structure important to predictive performance.

Figure~\ref{fig:sarma-results} shows the results of the SARMA experiment. Interestingly, while our procedure is able to identify seasonal lags that are both highly stable and close to the true model, \pkg{auto.arima} fails to recognise a seasonal component at all. This behaviour was seen across other simulated examples not shown here, and is perhaps an artifact of the AIC-based selection criterion. As well as this out-performance in the seasonal component, projection predictive inference is once more closer to the truth and more stable in its non-seasonal selection when compared to \pkg{auto.arima}.
\subsection{Predictive performance}\label{sec:predictive-performance}
Having shown that projection predictive model selection is considerably more stable than \pkg{auto.arima}, we presently show that this stability does not come at the expense of submodel predictive performance.

To this end, we use both projection predictive model selection  and \pkg{auto.arima} to identify parsimonious submodels for well-studied datasets curated by \citet{fpp2}. We then record the orders selected and the predictive performance of their MCMC fitted models as measured by LOO-CV elpd. Finally, we measure the difference between the models' elpds in an effort to ascertain the magnitude of difference between the two model's predictive performances.

We can understand difference in predictive performance between the two procedures through the elpd difference (diff.) column of tables~\ref{tab:nonseasonal-data} and \ref{tab:seasonal-data}. When this difference is positive, the mean elpd of the model selected by our procedure (Algorithm~\ref{algo}) was higher (better) than that selected by \pkg{auto.arima} (Algorithm~\ref{algo:autoarima}), and \textit{vice versa}. We embolden differences for which zero is not the approximate 90\% normal interval over the mean elpd difference estimate (the $1.64$ standard deviation range).\footnote{See the work of \citet{sivula2020uncertainty} for the properties of the sampling distribution of LOO-CV elpd differences.}
\begin{table*}[t!]
    \small
    \centering
    \begin{threeparttable}
    \newcolumntype{C}{>{$}c<{$}}
    \newcolumntype{L}{>{$}l<{$}}
    \newcolumntype{R}{>{$}r<{$}}
    \newcolumntype{z}{ @{}>{${}}l<{{}$}@{} }
    \newcolumntype{y}{ @{}>{${}}c<{{}$}@{} }
    \newcolumntype{x}{ @{}>{${}}r<{{}$}@{} }
    
    \begin{tabular}{l CCxyz CCxyz xyz}\toprule
     & \multicolumn{5}{c}{\textbf{ProjpredARMA}} & \multicolumn{5}{c}{\textbf{MCMC AutoARIMA}} & & &
    \\\cmidrule(lr){2-6}\cmidrule(lr){7-11}
    Data & p^\perp & q^\perp & \multicolumn{3}{c}{elpd $\pm$ s.e.} & p_{\auto} & q_{\auto} & \multicolumn{3}{c}{elpd $\pm$ s.e.} & \multicolumn{3}{c}{elpd diff. $\pm$ s.e.}\\\midrule
    A & 2 & 3 & -100 & \pm & 12 & 0 & 1 & -130 & \pm & 13 & \bm{30} & \bm{\pm} & \bm{7.2} \\
    B & 1 & 0 & 7.3 & \pm & 3.3 & 1 & 0 & 7.4 & \pm & 3.3 & 0.0 & \pm & 0.1 \\
    C & 2 & 3 & -108 & \pm & 7 & 2 & 1 & -110 & \pm & 7 & 1.4 & \pm & 1.0 \\
    D & 1 & 1 & -80 & \pm & 5 & 1 & 1 & -80 & \pm & 5 & 0.0 & \pm & 0.1 \\
    E & 3 & 0 & -608 & \pm & 25 & 0 & 2 & -613 & \pm & 25 & \bm{5.1} & \bm{\pm} & \bm{3.1} \\
    F & 2 & 0 & -110 & \pm & 7 & 0 & 1 & -109 & \pm & 6 & -1.0 & \pm & 1.4 \\
    G & 1 & 3 & 1.9 & \pm & 5.9 & 1 & 1 & 1.4 & \pm & 6.2 & 0.5 & \pm & 0.9 \\\bottomrule
    \end{tabular}
    
    \begin{tablenotes}
    \small
    \item \textbf{Dataset lookup:} A $=$ airline passengers ($n = 47$), B $=$ international visitors ($n = 36$), C $=$ Lake Huron bathymetry ($n = 98$), D $=$ insurance quotes ($n = 40$), E $=$ Ansett Airline passengers ($n = 269$), F $=$ maximum annual temperature ($n = 46$), G $=$ female murder rate ($n = 55$).
    \end{tablenotes}
    \end{threeparttable}
     
    \caption{Non-seasonal real world data. Predictive performance comparison of ProjpredARMA with MCMC AutoARIMA on non-seasonal data sets curated by \citet{fpp2}. For both procedures, we report the order of the suggested model (denoted $p^\perp$ and $q^\perp$ in the case of the projected orders, and $p_{\auto}$ and $q_{\auto}$ for \pkg{auto.arima}) as well as the models' elpd and elpd standard error (s.e.). We \textbf{embolden} the elpd differences such that zero is not included within the approximate 90\% normal interval over the mean elpd difference estimate (the $1.64$ standard deviation range). The number of observations in each dataset is reported by $n$ in the dataset lookup. Our results show that our procedure always produces models with predictive performance at least as good as \pkg{auto.arima}, and in three cases identified considerably better models.}
    \label{tab:nonseasonal-data}
\end{table*}
\begin{table*}[t!]
    \footnotesize
    \centering
    \begin{threeparttable}
    \newcolumntype{C}{>{$}c<{$}}
    \newcolumntype{L}{>{$}l<{$}}
    \newcolumntype{R}{>{$}r<{$}}
    \newcolumntype{z}{ @{}>{${}}l<{{}$}@{} }
    \newcolumntype{y}{ @{}>{${}}c<{{}$}@{} }
    \newcolumntype{x}{ @{}>{${}}r<{{}$}@{} }
    \addtolength{\tabcolsep}{-0.2em}
    \begin{tabular}{l CCCCxyz CCCCxyz xyz}\toprule
    & \multicolumn{7}{c}{\textbf{ProjpredARMA}} & \multicolumn{7}{c}{\textbf{MCMC AutoARIMA}} & 
    \\\cmidrule(lr){2-8}\cmidrule(lr){9-15}
    Data & p^\perp & q^\perp & P^\perp & Q^\perp & \multicolumn{3}{c}{elpd $\pm$ s.e.} & p_{\auto} & q_{\auto} & P_{\auto} & Q_{\auto} & \multicolumn{3}{c}{elpd $\pm$ s.e.} & \multicolumn{3}{c}{elpd diff. $\pm$ s.e.}\\\midrule
    H & 3 & 3 & 1 & 0 & -153 & \pm & 14 & 1 & 2 & 0 & 0 & -190 & \pm & 15 & \bm{38} & \bm{\pm} & \bm{9.3} \\
    I & 3 & 4 & 3 & 2 & 281 & \pm & 12 & 4 & 1 & 0 & 2 & 278 & \pm & 11 & 2.8 & \pm & 4.5 \\
    J & 3 & 4 & 1 & 0 & -273 & \pm & 19 & 1 & 1 & 0 & 1 & -282 & \pm & 22 & 8.7 & \pm & 8.4 \\
    K & 4 & 0 & 2 & 3 & -494 & \pm & 11 & 4 & 1 & 0 & 1 & -492 & \pm & 11 & -1.8 & \pm & 2.4\\
    L & 1 & 4 & 2 & 1 & -1470 & \pm & 32 & 1 & 2 & 2 & 2 & -1465 & \pm & 32 & -3.0 & \pm & 4.1\\\bottomrule
    \end{tabular}
    \begin{tablenotes}
    \small
    \item \textbf{Dataset lookup:} H $=$ Mona Loa $CO_2$ ($n = 468,\, s = 12$), I $=$ corticosteroid subsidy ($n = 204,\, s = 12$), J $=$ anti-diabetic drug subsidy ($n = 204,\, s = 12$), K $=$ equipment manufacturing ($n = 195,\, s = 12$), L $=$ daily electricity demand ($n = 365,\, s = 7$).
    \end{tablenotes}
    \end{threeparttable}
    
    \caption{Seasonal real world data. Predictive performance comparison of ProjpredARMA with MCMC AutoARIMA on seasonal data sets curated by \citet{fpp2}. For both procedures, we report the order of the suggested model as well as the models' elpd and elpd standard error (s.e.). We \textbf{embolden} the differences between elpds similarly to as in Table~\ref{tab:nonseasonal-data}. As before, we list the number of data points in each series under $n$ in the dataset lookup, as well as the identified seasonality $s$. These results show again that our procedure produces models at least as good as those chosen by \pkg{auto.arima} with regards to predictive performance, and twice out-performed the latter.}
    \label{tab:seasonal-data}
\end{table*}

We perform this experiment with data from the \pkg{fpp2} package \citep{fpp2} since they are well-studied, clean, and we can easily fit a reference model to them with the default priors previously discussed. If the data are deemed to be seasonal by \pkg{auto.arima}, we use the seasonal lag $s$ provided therein for our reference models. Similarly we use the same order of differencing (seasonal and non-seasonal) identified by \pkg{auto.arima} in both procedures for consistency. We disallow the possibility of drift in the models identified by \pkg{auto.arima} and concern ourselves only with inference in the stationary case.

We only consider data where the suggested differencing and seasonalities are inline with our prior knowledge, and appear to be reasonable upon closer inspection of the ACF and PACF plots. As is mentioned by \citet{autoarima}, we favour data where minimal differencing is required for improved predictive performance. Further, we use data such that the ACF and PACFs plots lead naturally to reference models through the Box-Jenkins approach \citep{boxjen:1970} that pass posterior checks when our default priors are used. It is worth mentioning that seasonality can also be directly ascertained from the ACF and PACF plots, but we prefer to use them only as sanity checks. Plots of the data used can be found in Appendix~\ref{sec:appendix-data}.

It is important to remember that the results we achieve are entirely dependent on the reference model used in the projection procedure. Indeed, this allows the statistician to encode prior beliefs into their model selection procedure through the reference model, a luxury unavailable in \pkg{auto.arima}. 

In Table~\ref{tab:nonseasonal-data} we present the predictive performance of the models selected by both procedures across a set of non-seasonal \pkg{fpp2} datasets. We find that the difference between the mean elpds is almost always positive (meaning that the mean elpd of the ProjpredARMA-achieved submodel is almost always higher than that of the submodel suggested by MCMC AutoARIMA), apart from in datasets A and E (airline passenger data and Ansett Airline data) where we find that the difference between elpds is skewed more than 1.64 standard deviations in our procedure's favour. We thus find that our procedure is able to achieve increased stability without sacrificing predictive performance, and in some cases even out-performs \pkg{auto.arima}.

Moving to seasonal experiments, we tabulate the results in Table~\ref{tab:seasonal-data}. In one of the five examples (Mona Loa $CO_2$), projection predictive model selection was able to find a submodel out-performing that selected by \pkg{auto.arima}. In the remaining instances, the mean difference between elpds remains mostly positive (in our favour). Again, we find that our improved stability does not come at the cost of predictive performance, and indeed some improvement in the latter is also felt.
\subsection{Alternative projections}\label{sec:bad-proj}
\begin{figure*}[t!]
    \centering
    \includegraphics{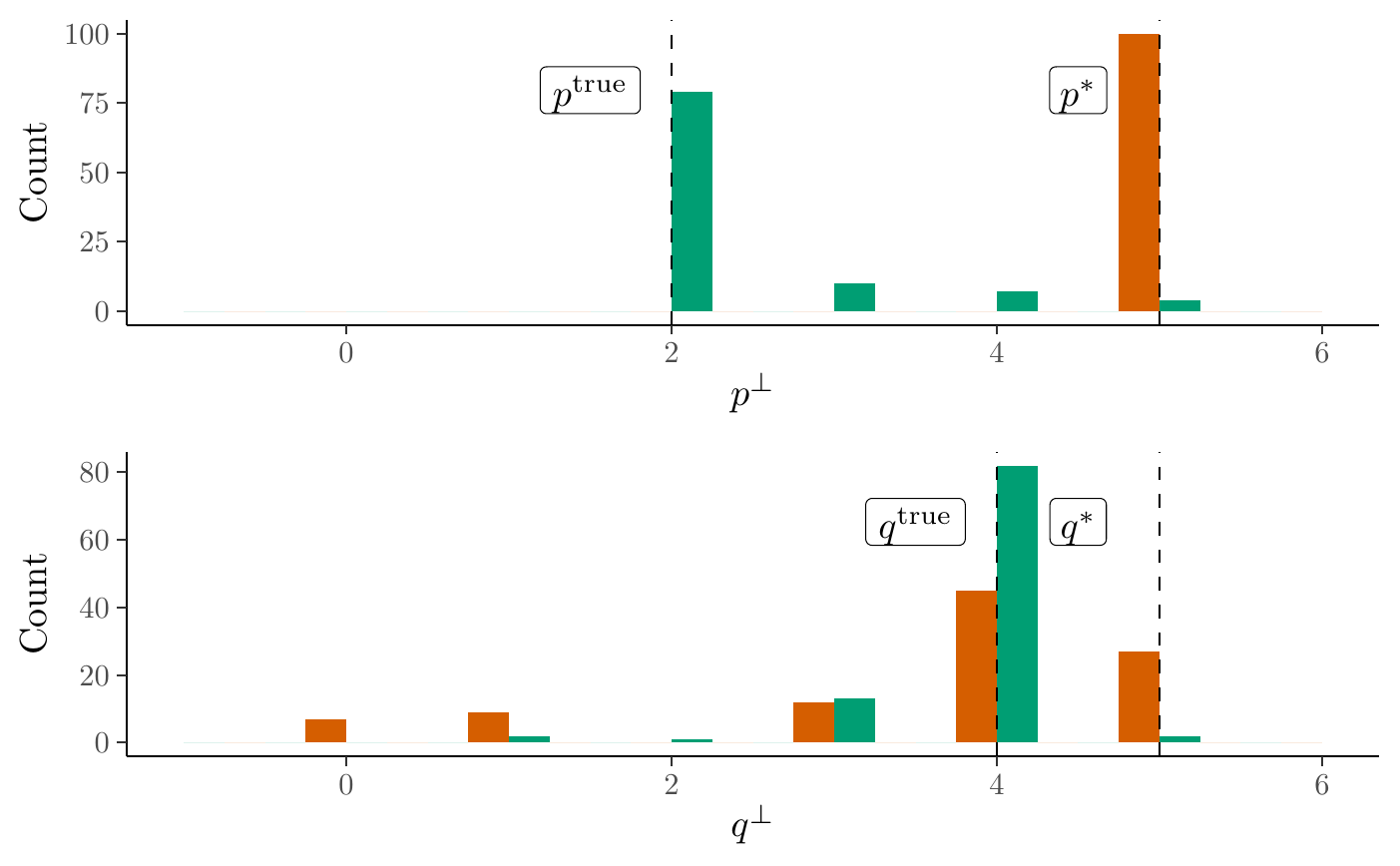}
    \caption{Simulated data. A frequency plot of the selected model orders under the procedure defined in Algorithm~\ref{algo} in \textcolor{projpredcolour}{green} and the modification to it described in Section~\ref{sec:bad-proj} in \textcolor{badcolour}{red}. The $100$ series were generated from an $\arma(2,4)$ process. The reference model parameters used, $p^\ast$ and $q^\ast$, and the true data-generating parameters $p^\text{true}$ and $q^\text{true}$ shown by the dotted black lines. We find that our procedure as defined in Algorithm~\ref{algo} correctly handles auto-regressive and moving-average information, and that the modification described in Section~\ref{sec:bad-proj} risks over-selecting auto-regressive components and under-selecting moving-average components.}
    \label{fig:bad-proj}
\end{figure*}
\begin{figure*}[t!]
    \centering
    \includegraphics{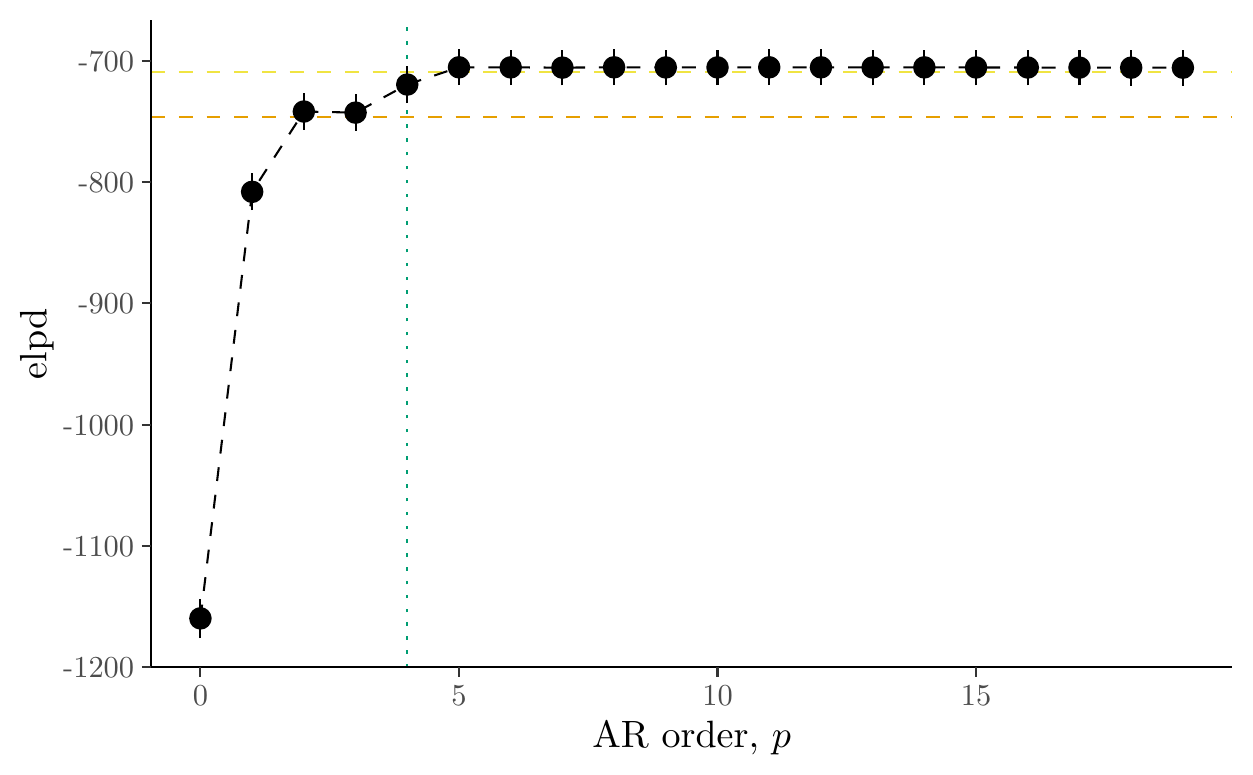}
    \caption{Simulated data. The normal-approximation $68\%$ elpd confidence intervals of $\ar(p)$ models produced by projection predictive inference are shown in the black point ranges. The predictive performance of the reference model is seen in the horizontal dashed \textcolor{refcolour}{yellow} line, and model size selected by our search heuristic from Section~\ref{sec:select} is shown in the vertical dotted \textcolor{projpredcolour}{green} line. The predictive performance of the $\arma(3,3)$ model found by \pkg{auto.arima} is seen in the horizontal dashed \textcolor{autocolour}{orange} line.}
    \label{fig:ar-infinite}
\end{figure*}
As was previously stated in Section~\ref{sec:arma-to-ar-proj}, it is necessary to separate the reference model into its AR and MA components before performing projection predictive inference. Indeed, in Figure~\ref{fig:bad-proj} we show that when we do not seperate these components in the reference model, we are liable to vastly over-select the size of the restricted AR model and slightly under-select the size of the restricted MA model, since we project too much information onto this restricted AR component. This results in submodels requiring larger AR components, which in turn produce noisier residuals and smaller MA components. We show this by performing the modification to our procedure discussed in Section~\ref{sec:arma-to-ar-proj} to $100$ simulated series from an $\arma(2,4)$ process. For comparison, we also show the selection frequencies of our original procedure in Algorithm~\ref{algo} and the true process values, confirming our suspicions that such modifications incorrectly incorporate information from the reference model in submodels.

There are instances, as we have previously discussed, where a projection directly from an ARMA model to an AR model can reduce computational burden in inference. What is then of interest to us is to understand how large the AR order must be to replicate the predictive performance of the $\ar(\infty)$ model implied by the non-zero MA component of the ARMA model. To demonstrate this, we presently sample a series from an $\arma(2, 3)$ model, and fit to it a reference $\arma(5,5)$ model. We then perform the projection from this $\arma(5,5)$ onto $\ar(p),\,p=1,\dotsc,20$, cross-validating the models' performances, and show the results in Figure~\ref{fig:ar-infinite}. We conclude that our $\arma(5,5)$ model's predictive performance can be replicated by as small a model as an $\ar(4)$. Not only this, but when we use \pkg{auto.arima} to identify a more parsimonious model for these data, we are returned an $\arma(3,3)$. This model is neither easy to infer, nor does it match our reference model -- which we are able to do by construction. Thus we find that this $\arma$ to $\ar$ projection can not only out-perform \pkg{auto.arima}, but can do so at much reduce cost.
\subsection{Robustness to noise}
\begin{figure*}[t!]
    \centering
    \includegraphics{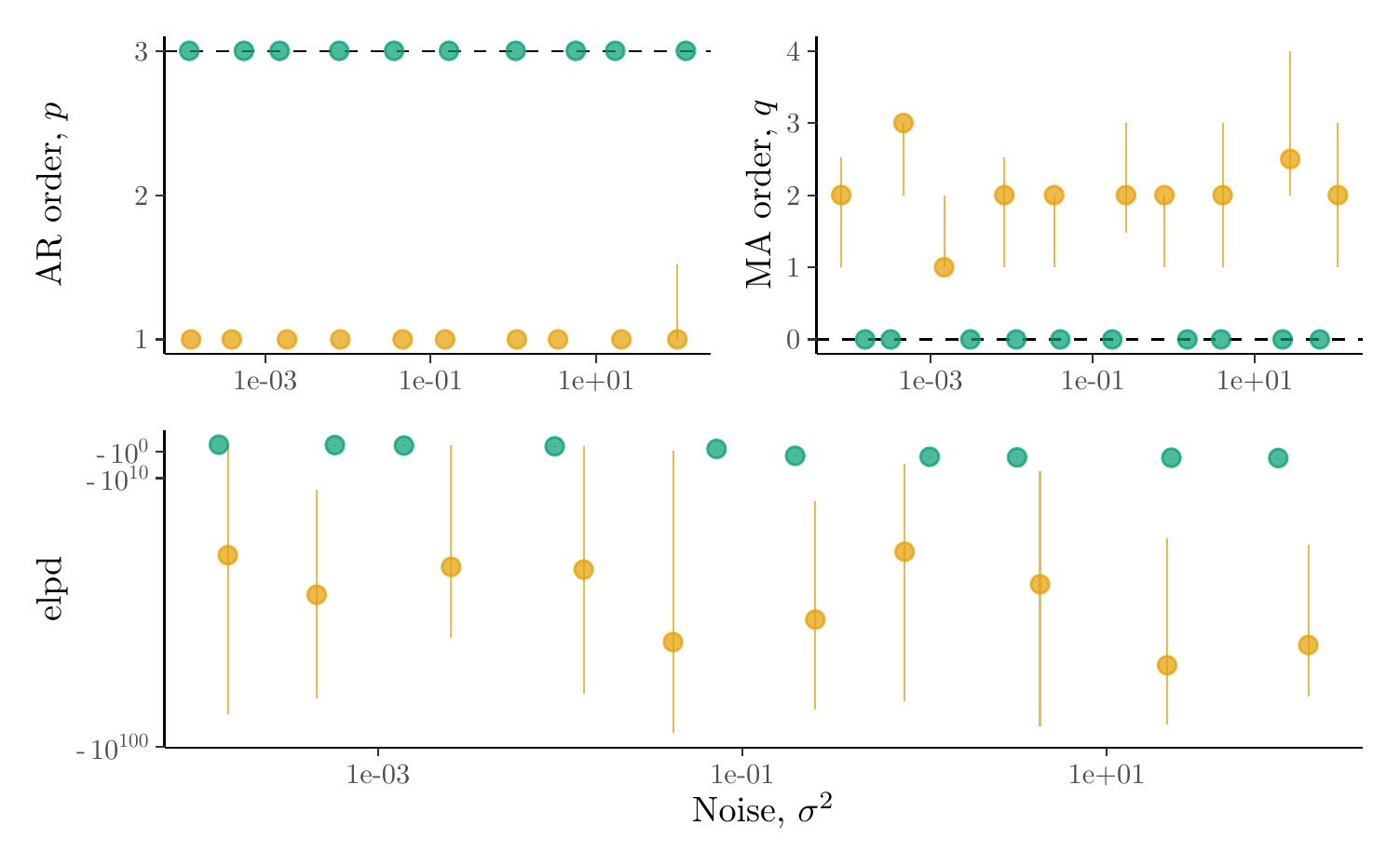}
    \caption{Simulated data. In the bottom axes are shown the elpds of models selected by projection predictive model selection in \textcolor{projpredcolour}{green} and by \pkg{auto.arima} in \textcolor{autocolour}{orange} from a simulated $\ar(3)$ model, corrupted with different levels of noise, $\sigma^2$, seen along the $x$-axis. The one standard error interval of observed mean elpds is also shown, and we note the scale of the axis. The top axes show the size of AR component identified by the the submodels at each value of $\sigma^2$ (in the top left panel), and the MA order identified likewise (in the top right). We find that across all levels of noise, \pkg{auto.arima} fails to recognise the true temporal structure while projection predictive inference chooses the same (and true) model orders independently of noise. Further, we see that \pkg{auto.arima} produces models with significantly worse poor predictive performance than our procedure.}
    \label{fig:snr}
\end{figure*}
A major advantage to the Bayesian workflow is the ability to manage uncertainty in data. It is then interesting for us to understand how our two competing procedures behave when different levels of noise are injected into the same underlying process.

We produce data from an $\ar(3)$ model with different levels of noise $\sigma$ injected. Formally, we sample from 
\begin{IEEEeqnarray}{rcl}
y_t \;&=\;& \phi_1y_{t-1} + \phi_2y_{t-2} + \phi_3y_{t-3} + \varepsilon_t \nonumber \\
\varepsilon_t \;&\sim\;&\normal(0,\sigma^2). \nonumber
\end{IEEEeqnarray}
We then apply both projection predictive model selection and \pkg{auto.arima} procedures to identify submodels for each of the series, compute their predictive performance, and show the results in Figure~\ref{fig:snr}.

We note three important results from this experiment convincing us of the robustness to noise afforded by our proposed procedure. First we find that projection predictive inference produces models with elpd at least as good as those identified by \pkg{auto.arima} under all noise levels. 

Second we find that \pkg{auto.arima} is liable to produce ill-performing models, seen in the instances of very low and uncertain elpd compared to projection predictive inference. Uncertainty in the selected orders is additionally much larger compared to our approach, in line with the results in Section~\ref{sec:stability}.  It is important to note the scale of the elpd plot; as a heuristic, we might consider that two elpd values with a difference of less that $4$ are essentially equivalent, whereas in this plot we show differences of several orders of magnitude indicating an infeasible model. These results suggest that some of the models identified by \pkg{auto.arima} may be misspecified given the data.

Finally, projection predictive model selection is exceptionally stable in model size, which does not vary greatly among the different levels of $\sigma^2$, whereas \pkg{auto.arima} exhibits large variability when selecting the moving-average component. Further, we find that projection predictive inference identifies the true model under all noise regimes.

Such behaviour is expected from projection predictive inference. Indeed, \citet{piironen2016} discuss how the use of a reference model in model selection (and more so in projection predictive model selcetion) is able to filter out noise from the data. We are thus confident in projection predictive inference's robustness to noisy data.
\subsection{The effect of distant lags on submodel selection}
\begin{figure*}[t!]
    \centering
    \includegraphics{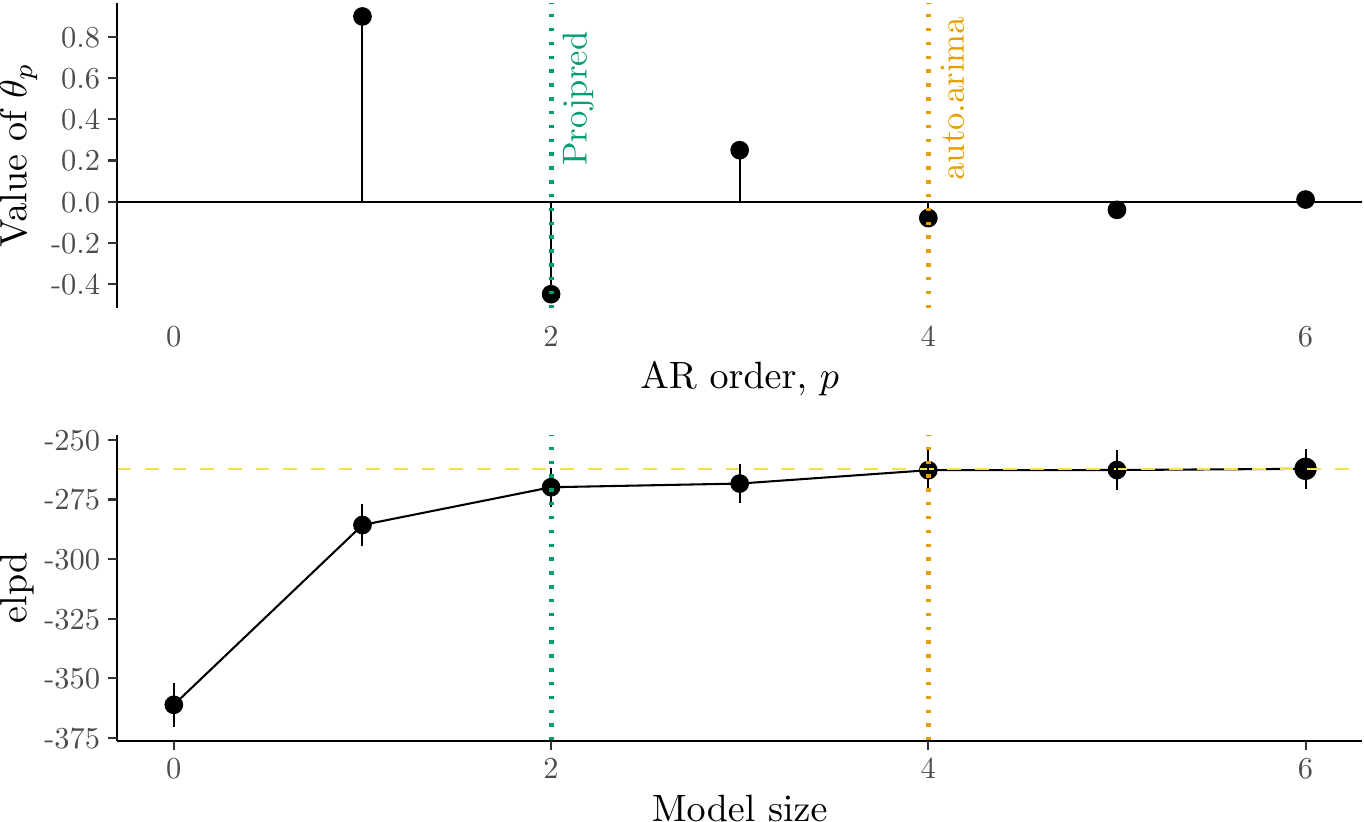}
    \caption{Simulated data. The top axes show the values of the true data generating $\ar(6)$ process parameters $\theta_p$ at each lag $p$, below which are the normal-approximation $68\%$ elpd confidence intervals of models selected by projection predictive inference. The predictive performance of the reference model is seen in the dashed \textcolor{refcolour}{yellow} line. The model size selected by our search heuristic from Section~\ref{sec:select} is shown in the dotted \textcolor{projpredcolour}{green} line, while that of \pkg{auto.arima} is indicated by the dashed \textcolor{autocolour}{orange} line. This plot shows our model selection heuristic behaviour's in a simple case where it finds distant lags to have predictive effect close to zero, while \pkg{auto.arima} has a tendency to select slightly larger models for negligible performance gain.}
    \label{fig:diminishing}
\end{figure*}
As has been previously discussed, the primary aim of projection predictive model selection is to find the smallest submodel such that its posterior predictive performance is not significantly different to that of a reference model. In the case of additive models, the addition of an additive component can only ever improve the posterior predictive performance. Thus, if we have many distant lags with low influence in some true data generating process, then our procedure will select only the first few necessary to encapsulate the expressiveness of the reference model and allow it to generalise well to new data.

One such model is given in Figure~\ref{fig:diminishing}, in which we have an $\ar(6)$ where the values of $\theta_p$ decrease with $p$ until lag six, after which all lags are equal to exactly zero. We simulate $N=200$ data points from this process and apply our projection predictive model selection procedure with an oracle providing an $\ar(6)$ reference model. Consequently, we identify an $\ar(2)$ model as the smallest submodel with comparable posterior predictive performance, since the mean reference model elpd lies within one standard error from the mean submodel elpd (the normal-approximation $68\%$ interval as previously described). However, using \pkg{auto.arima} to identify a parsimonious submodel, again with an oracle providing information that the process is a stationary non-seasonal auto-regressive model, identifies an $\ar(4)$ model. This over-fitting of weak-effect covariates is common in AIC-based validation, while projection predictive model selection is able to avoid it by prioritising predictive performance over model complexity and is able to lean on a reference model in order to do so.
\subsection{Scalability}
We have dealt with examples where the number of observations has been fewer that $500$. 
The computation time of Algorithm~\ref{algo} increases linearly with the number of data observations. Indeed, even for very large datasets we can still perform our procedure in reasonable time. It is worth noting that most of this computational time is spent fitting the AR and MA components with MCMC, and as a result is dependent on the priors used and machinery chosen. Naturally, given this Bayesian treatment of order identification our procedure remains computationally intensive when compared to \pkg{auto.arima}.
\section{Conclusion}\label{sec:conclusion}
We have motivated an extension of projection predictive model selection to probabilistic ARMA models and thence developed a novel two-stage Bayesian order identification procedure. Our procedure was shown through simulated and real-data experiments to:
\begin{enumerate}[nosep]
    \item be stable in model selection, and robust to instances of noisy data and complex data-generating models;
    \item produce submodels with predictive performance at least as good as \pkg{auto.arima};
    \item scale well with increased data size, although it remains computationally expensive when compared to \pkg{auto.arima}.
\end{enumerate}
Importantly, we have shown how the original idea of \citet{goutis1998} and \citet{dupuis2003} can be abstracted beyond the realm of generalised linear models and towards time series applications, wherein parsimony and model structure is a key concern. In doing so, we have motivated a robust and efficient ARMA order identification procedure from an information theory perspective.
\section{Discussion}
Model selection is often motivated as a remedy to over-fitting. However, in a Bayesian regime the statistician is afforded the luxury of explicating their prior beliefs. The problem of over-fitting can then be mitigated at least in part with the use of sensible priors. Instead, we propose our procedure for use in one of three candidate situations:
\begin{enumerate}[nosep]
    \item we have a rich model which is good for prediction, but would like to reduce its size to reduce the computational burden or to improve robustness to changes in the data-generating distribution;
    \item we would like to identify predictive submodels to gain a better understanding of important temporal correlation structures;
    \item we have an ARMA model with moving-average components but would like to convert it to a purely auto-regressive model, or more generally we would like to investigate models with similar predictive performance but different structures.
\end{enumerate}
Importantly, we do not advocate for model selection in a Bayesian setting for its own sake. Rather we believe that model selection within a reference model paradigm can provide the statistician improved interpretation of the underlying data-generating process, and the opportunity to reduce computational cost without sacrificing predictive performance. As such, the creation of the reference model and the use of priors therein is of critical importance to a good analysis.

The choice of reference model is not unambiguous in general, and the results of our procedure may vary with the ability of the prior to distinguish noise from signal in the data \citep{kohns2022flexible}.

The procedure presented in this paper has dealt with the ARMA model given its prevalence in literature and empirically proven strength in practice. We believe that other time series models, notably state space and Gaussian process time series models could both also benefit from projection predictive model selection. Indeed \citet{projpred-gp} have previously shown that it is able to deal with Gaussian process model selection, and \citet{projpredgam} have likewise shown its extension to generalised additive models.

We have also discussed how there may arise situations in which directly using standard sparsifying priors such as the horseshoe and spike-and-slab priors \citep{horseshoe, spikeslab} for ARMA variable selection may result in invalid or unintuitive models. Further investigation into the sparsifying time series priors, possibly similar to the R2D2 prior of \citet{r2d2}, the $L_1$ ball prior of \citet{xu2020}, or fused lasso by \citet{casella2010penalized} would surely benefit the field.
\subsection*{Acknowledgements}
We acknowledge the computational resources provided by the Aalto Science-IT project. This paper was partially funced by the Research Council of Finland Flagship programme: Finnish Center for Artificial Intelligence, and Research Council of Finland project ``Safe iterative model building'' (340721).
\bibliography{projpred-arma.bib}
%
% appendix
\newpage
\onecolumn
\begin{appendices}
\section{Datasets}\label{sec:appendix-data}
\begin{figure*}[t!]
    \centering
    \includegraphics{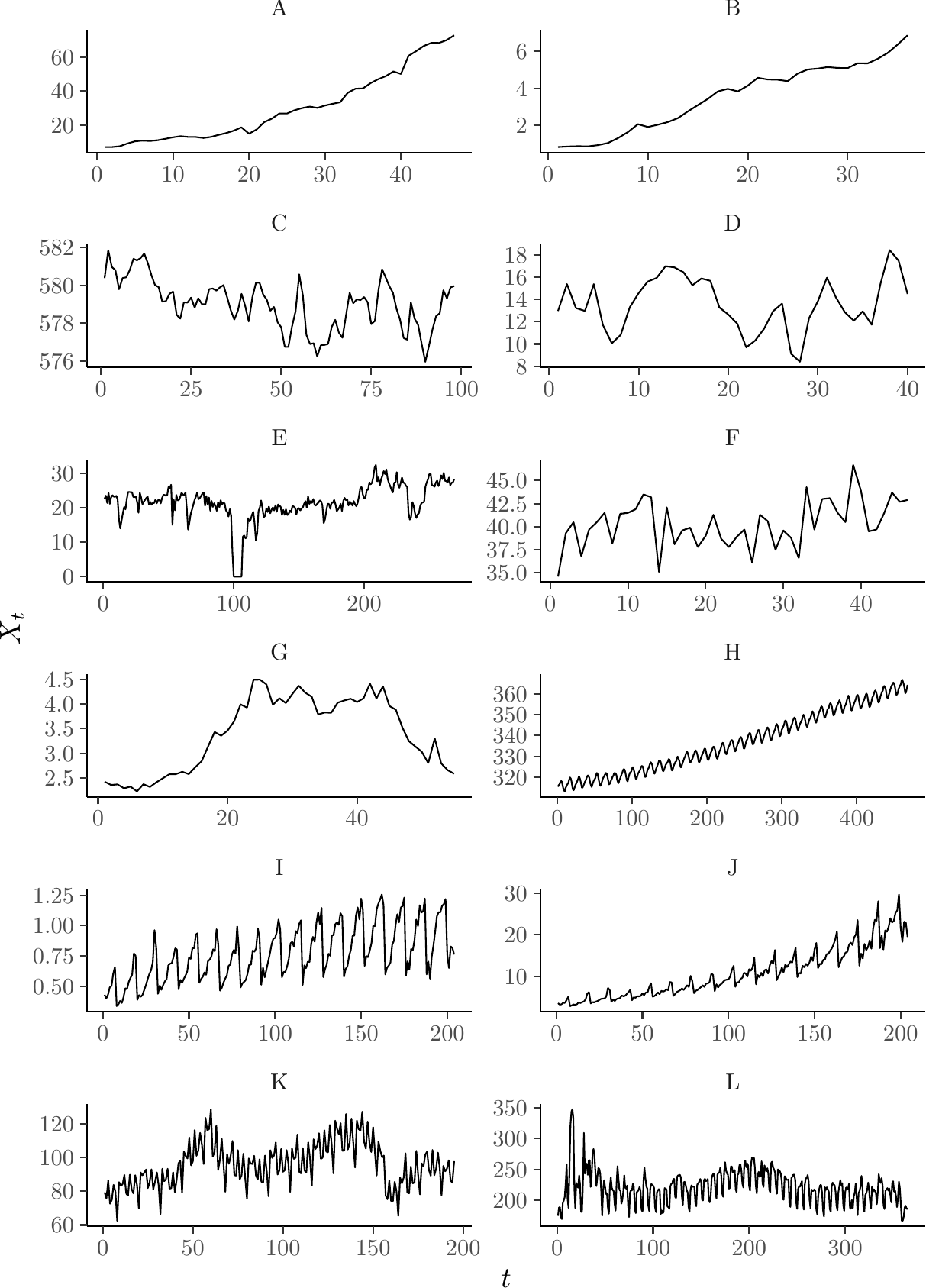}
    \caption{Datasets used in Section~\ref{sec:predictive-performance}. In the lookup below, we also note the length of the time series by $n$ and the seasonality identified by \pkg{auto.arima} by $s$.\\
    \hspace{\parindent} \textbf{Dataset lookup:} A $=$ airline passengers ($n = 47$), B $=$ international visitors ($n = 36$), C $=$ Lake Huron bathymetry ($n = 98$), D $=$ insurance quotes ($n = 40$), E $=$ Ansett Airline passengers ($n = 269$), F $=$ maximum annual temperature ($n = 46$), G $=$ female murder rate ($n = 55$), H $=$ Mona Loa $CO_2$ ($n = 468,\, s = 12$), I $=$ corticosteroid subsidy ($n = 204,\, s = 12$), J $=$ anti-diabetic drug subsidy ($n = 204,\, s = 12$), K $=$ equipment manufacturing ($n = 195,\, s = 12$), L $=$ daily electricity demand ($n = 365,\, s = 7$).}
    \label{fig:datasets}
\end{figure*}

\begin{figure*}[t!]
    \centering
    \includegraphics{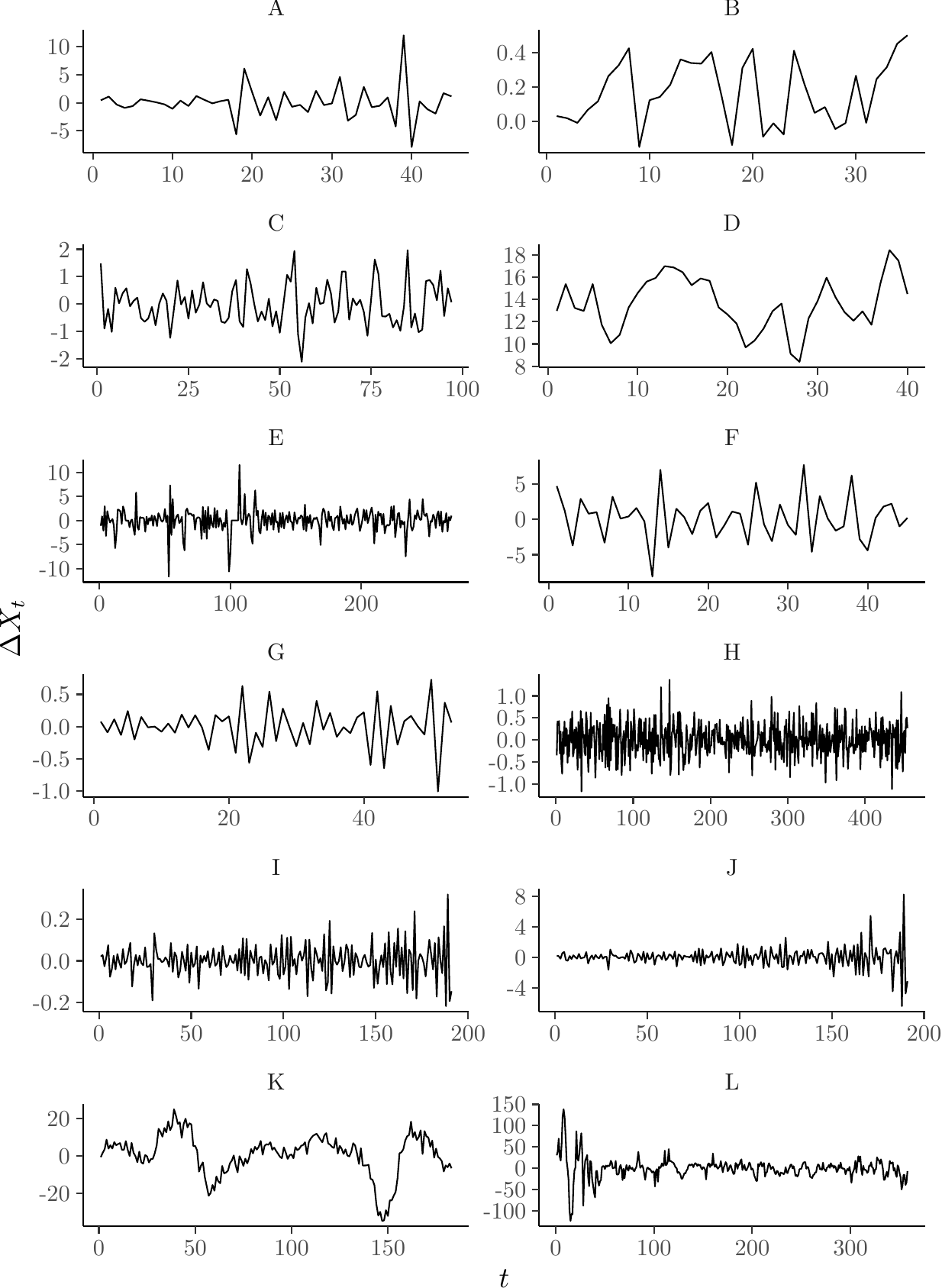}
    \caption{Differenced datasets (denoted $\Delta y_t$) used in Section~\ref{sec:predictive-performance}. We record the order of non-seasonal and seasonal differencing denoted $d$ and $D$ respectively.\\
    \hspace{\parindent} \textbf{Dataset lookup:} A $=$ airline passengers ($d = 2$), B $=$ international visitors ($d = 1$), C $=$ Lake Huron bathymetry ($d = 1$), D $=$ insurance quotes ($d = 0$), E $=$ Ansett Airline passengers ($d = 1$), F $=$ maximum annual temperature ($d = 1$), G $=$ female murder rate ($d = 2$), H $=$ Mona Loa $CO_2$ ($d = 1,\, D =1$), I $=$ corticosteroid subsidy ($d = 1,\, D =1$), J $=$ anti-diabetic drug subsidy ($d = 1,\, D =1$), K $=$ equipment manufacturing ($d = 0,\, D =1$), L $=$ daily electricity demand ($d = 0,\, D =1$).}
    \label{fig:datasets-diff}
\end{figure*}

We present below the raw data used in the experiments of Section~\ref{sec:predictive-performance} in Figure~\ref{fig:datasets}, and in Figure~\ref{fig:datasets-diff} we show them after applying the differencing suggested by \pkg{auto.arima}.
\end{appendices}
\end{document}